\documentclass{INTERSPEECH2023}

\interspeechcameraready

\usepackage[utf8]{inputenc} 
\usepackage[T1]{fontenc}    
\usepackage{hyperref}       
\usepackage{url}            
\usepackage{booktabs,arydshln}       
\usepackage{amsfonts}       
\usepackage{nicefrac}       
\usepackage{microtype}      
\usepackage{graphicx}
\usepackage{float}
\usepackage{multirow}
\usepackage{amssymb}
\usepackage{amsmath,amscd}
\usepackage{amsthm}
\usepackage{cancel}     
\usepackage{centernot}  
\usepackage{graphicx}   
\usepackage{soul}       
\usepackage{color}      
\usepackage{enumitem}   
\usepackage{commath}    
\usepackage{subcaption} 
\usepackage{xcolor,colortbl}
\usepackage{CJK}

\definecolor{lightergray}{gray}{0.85}
\newcolumntype{a}{>{\columncolor{lightergray}}c}
\newcolumntype{b}{>{\columncolor{lightergray}}l}

\newlength{\bibitemsep}
\newlength{\bibparskip}
\let\oldthebibliography\thebibliography
\renewcommand\thebibliography[1]{%
  \oldthebibliography{#1}%
  \setlength{\parskip}{\bibitemsep}%
  \setlength{\itemsep}{\bibparskip}%
}

\title{Syllable Discovery and Cross-Lingual Generalization in\\ 
 a Visually Grounded, Self-Supervised Speech Model}
\name{Puyuan Peng$^1$, Shang-Wen Li$^2$, Okko R{\"a}s{\"a}nen$^3$, Abdelrahman Mohamed$^4$, David Harwath$^1$}
\address{
  $^1$Department of Computer Science, The University of Texas at Austin, USA\\
  $^2$Meta AI, USA, 
  $^3$Unit of Computing Sciences, Tampere University, Finland,
  $^4$Rembrand, USA}
\email{pyp@utexas.edu}

\begin{document}
\begin{CJK*}{UTF8}{gbsn}

\maketitle
 
\begin{abstract}
In this paper, we show that representations capturing syllabic units emerge when training a self-supervised speech model with a visually-grounded training objective. We demonstrate that a nearly identical model architecture (HuBERT) trained with a masked language modeling loss does not exhibit this same ability, suggesting that the visual grounding objective is responsible for the emergence of this phenomenon. We propose the use of a minimum cut algorithm to automatically predict syllable boundaries in speech, followed by a 2-stage clustering method to group identical syllables together. We show that our model not only outperforms a state-of-the-art syllabic segmentation method on the language it was trained on (English), but also generalizes in a zero-shot fashion to Estonian. Finally, we show that the same model is capable of zero-shot generalization for a word segmentation task on 4 other languages from the Zerospeech Challenge, in some cases beating the previous state-of-the-art.\footnote{Code \& Model: \href{https://github.com/jasonppy/syllable-discovery}{https://github.com/jasonppy/syllable-discovery}.}
\end{abstract}

\section{Introduction}
Traditionally, automatic speech recognition, speech synthesis, and spoken language understanding tasks have relied on supervised learning and the assumption that ground-truth text transcriptions of the training speech are available. Such transcriptions are costly to collect and represent a major hurdle in developing speech recognition and related technologies that can serve the thousands of languages around the world.

Recently the speech community has made tremendous progress developing self-supervised models that can learn powerful representations of the speech signal by being pre-trained on untranscribed speech data. After pre-training the models can be fine-tuned on a small amount of transcribed data to achieve impressive performance on a variety of tasks~\cite{Baevski2020wav2vec2A,hsu2021hubert,Chung2021w2vBERTCC,Chen2021WavLMLS,Mohamed2022SelfSupervisedSR}. Furthermore, the representations learned by these models can be clustered into discrete speech units that have been shown to be strongly correlated with words and phones~\cite{Harwath2019LearningHD,Peng2022WordDI}. These units can be used to tokenize speech into a pseudo-text sequence, which can be used as a drop-in replacement for a text transcription in a wide variety of downstream tasks, giving rise to a new genre of ``textless'' speech processing research~\cite{lakhotia-etal-2021-generative,li2022textless,Nguyen2022GenerativeSD,lin2022dual}.

Because of the emergent nature of these units, it is not yet understood how to control what type of linguistic structure (e.g. phones, syllables, words) they will capture. It has been shown that the representations of self-supervised speech models tend to correlate with lower-level structure such as phones at lower model layers, and higher-level structure such as words at higher model layers~\cite{Harwath2019LearningHD,pasad2022comparative}. However, it has also been demonstrated that the model's training objective strongly influences the nature of these representations. Training the model to perform cross-modal grounding of speech to contextually-relevant visual images has been shown to dramatically increase the model's word learning capability over a masked language modeling objective, even when the model architecture is held nearly constant~\cite{Peng2022WordDI}.

In this paper, we build on~\cite{Peng2022WordDI} and demonstrate that multimodal self-supervision simultaneously results in the emergence of word-like and syllable-like representations within the same model. While~\cite{Peng2022WordDI} showed that word-like units are encoded by the Transformer's attention heads, we show that syllabic structure emerges within the embeddings of the token sequence itself. We propose the use of a minimum cut segmentation algorithm to derive syllable boundaries from these features, outperforming a state-of-the-art method for unsupervised syllabic segmentation. We then show that these segments can be clustered across a speech corpus to perform syllable discovery, enabling tokenization of the speech signal at the level of syllable-like units. Finally, we also show surprising results where our model trained only on English speech is able to perform zero-shot segmentation of syllables on another language (Estonian) and words in multiple non-English languages, in several cases outperforming the state-of-the-art models on the Zerospeech challenge~\cite{Dunbar2022SelfSupervisedLL}. 

\section{Related Work}
Besides the aforementioned work on self-supervised and textless speech processing, our work is also related to spoken term discovery and visually grounded speech processing. 

Spoken term discovery - inferring the temporal boundary and identity of words and short phrases from untranscribed speech audio data - has been an important research direction in Zero-resource speech processing~\cite{Dunbar2022SelfSupervisedLL}. The earliest work that tackles spoken term discovery date back to at least the segmental dynamic programming algorithm proposed by Park and Glass~\cite{park08}. Since then, numerous other approaches have been proposed. \cite{lee-etal-2015-unsupervised,Taniguchi2015NonparametricBD} developed Bayesian models for hierarchical phoneme and word discovery. 
Based on the fact that syllables are organized around particularly sonorous speech sounds, \cite{rasanen18} developed sonority fluctuation-based method for syllabic segmentation.
Other works model word directly either via an iterative segmentating-clustering approach~\cite{Kamper2017AnES}, or reinforcement learning~\cite{Wang2018SegmentalAW}. Self-supervised learning has also been considered for end-to-end phoneme and word segmentation~\cite{Bhati2021SegmentalCP,Cuervo2021ContrastivePS}.
Mostly recently, 
Algayres et al.~\cite{Algayres2022DPParseFW} identified the key issues in applying text-based models for speech segmentation, and proposed the DP-Parse algorithm which uses instance lexicon to mitigate clustering error. Herman~\cite{Kamper2022WordSO} applied vector quantization for phoneme-like unit discovery, and then ran a dynamic programming algorithm on the discovered units for word segmentation. 

Visually grounded speech (VGS) processing~\cite{Chrupaa2021VisuallyGM} generalizes the idea of self-supervised learning to multimodal (visual) data and learns speech representations by associating speech audio with contextually-relevant visual input. VGS usually leverages image-speech~\cite{Gabriel2014LearningWF,hawath2016} or video-speech~\cite{Rouditchenko2020AVLnetLA,Nikolaus2022LearningEW} paired data.  
In practice, besides speech-image retrieval and alignment~\cite{Kamper2017SemanticSR,Sanabria2021TalkDW,fastvgs,Shih2022SpeechCLIPIS,Khorrami2021EvaluationOA,Harwath2018JointlyDV}, VGS models has also be shown to achieves competitive performance keyword spotting~\cite{Olaleye2023VisuallyGK}, query-by-example research~\cite{Kamper2019SemanticQS}, and varies tasks in the SUPERB benchmark~\cite{Yang2021SUPERBSP,fastvgs+}. The study of linguistic information learned in VGS models has been attracting increasing attention. In particular, researchers has measured the phonetic, syllabic, and lexical information in VGS models~\cite{Alishahi2017EncodingOP,Rsnen2019ACM,Harwath2019LearningHD,Havard2019ModelsOV,Havard2019WordRC,Peng2022WordDI,Khorrami2021CanPS}. In addition to~\cite{Peng2022WordDI}  which we build our work on, \cite{Khorrami2021CanPS} is the most relevant to ours where they studied the emergence of phonetic, syllabic, and lexical information in different layers of CNN-based VGS models. Our work is different from their in that none of the modules of our model receives textual supervision, while their image encoder is pre-trained on Imagenet classification~\cite{Russakovsky2014ImageNetLS}. In addition, we show the emergence of hierarchical linguistic information in the non-hierarchical Transformer model, while they use hierarchical CNN models.

\section{Technical Approach}
\begin{figure}
  \centering
      \includegraphics[width=1\columnwidth]{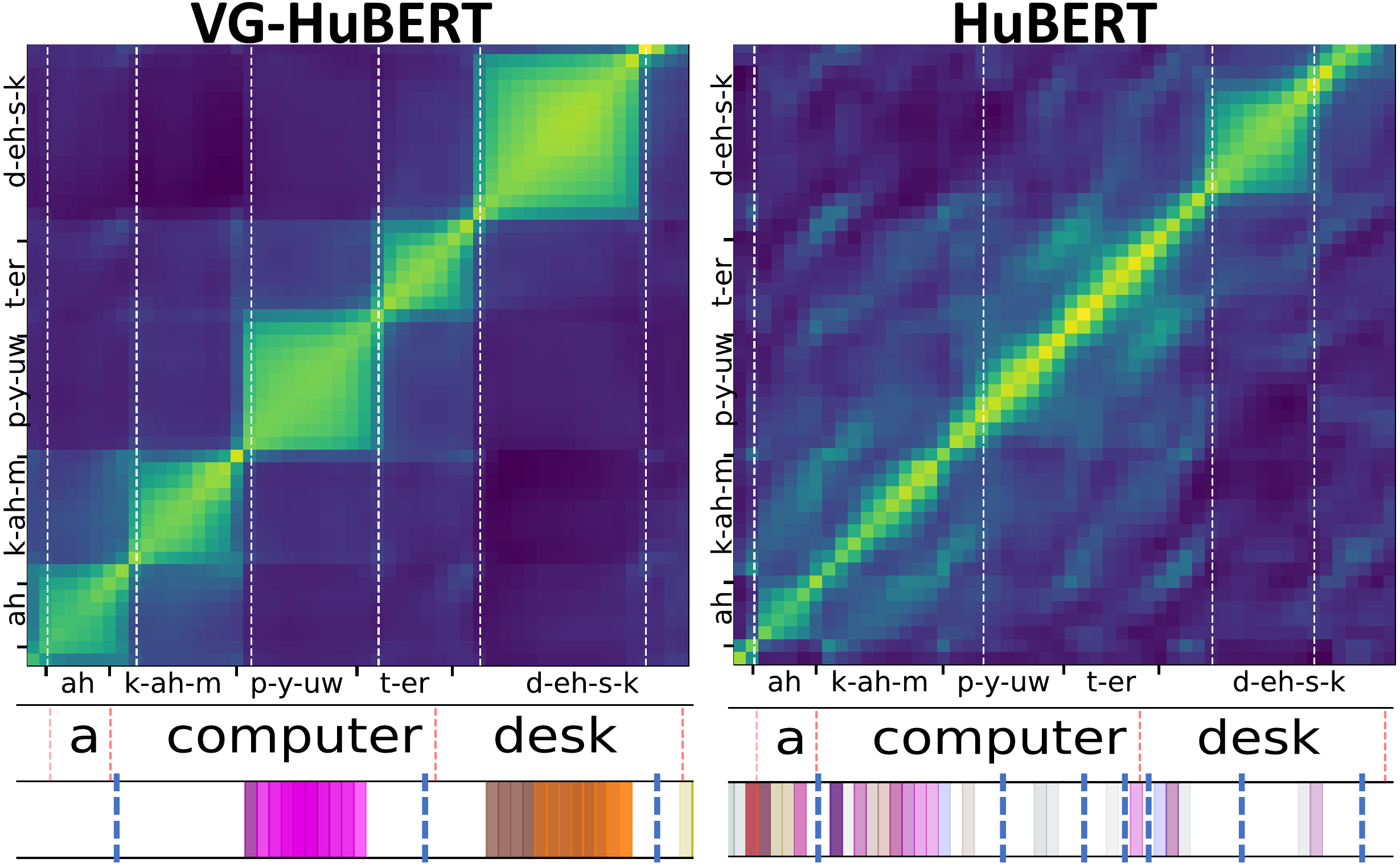} 
      \vspace{-5mm}
      \caption{Visualization of feature self-similarity matrix (upper) and the attention (lower) in VG-HuBERT and HuBERT. The vertical white dotted lines are generated by minCutMerge, and vertical blue dotted lines are generated by taking the midpoint of boundaries of adjacent attention segments}\label{fig:minCut_clsAttn}
      \vspace{-5mm}
\end{figure}

VG-HuBERT~\cite{Peng2022WordDI} is a self-supervised dual-encoder model trained using a contrastive loss to match speech waveforms with the images they describe. Although VG-HuBERT is not trained with any textual supervision, the model has been shown to exhibit strong word discovery capabilities~\cite{Peng2022WordDI}. Specifically, its CLS token places concentrated chunks of attention weight on word segments in input utterances (see lower left subfigure of figure~\ref{fig:minCut_clsAttn} for an example). 
Our motivating hypothesis is that VG-HuBERT's word discovery ability is predicated on its ability to also discover sub-word units at earlier layers. To probe this we first extract a sequence of frame embeddings from some layer of the model given an input waveform, $\mathbf{C} \in \mathbb{R}^{T\times D}$, ($T$ is number of speech frames, $D$ is the feature dimension). Next, we then calculate the feature self-similarity matrix as
$
\text{featSSM} := \mathbf{C}\mathbf{C}^{\intercal}
$. We normalize featSSM by subtracting smallest element of the matrix from all elements to insure that all frame-pair similarity scores are non-negative.
Figure~\ref{fig:minCut_clsAttn} shows an example of featSSM, where green color denotes high similarity and blue denotes low similarity. We see a clear block diagonal structure in VG-HuBERT's featSSM, where each block corresponds to a syllable. In HuBERT's featSSM, however, the block structure hardly exists. Based on the different patterns we see between the feature self-similarity matrix and the CLS attention, we hypothesize that visually grounded training leads to the emergence of syllable identity being encoded in VG-HuBERT's features, and the CLS token attending to these features to infer the presence of words. To quantitatively study the syllable discovery phenomenon, we adopt the normalized minimum cut algorithm~\cite{mincut1997,mincut2006,minCutMerge2012} to automatically segment the blocks in featSSM, and use the block boundaries to predict syllable boundaries.

\textbf{A min-cut segmentation algorithm for featSSM.} We define a fully-connected, undirected graph $G(V,E)$ for every speech utterance. Set $V$ consists of all speech frames as nodes; Set $E$ consists of edges, where the edge weight $w(u,v)$ is defined as the similarity score corresponding to nodes $u$ and $v$. Segmenting the blocks in featSSM means partitioning the corresponding graph $G(V,E)$ into disjoint sets $A_1, A_2, \cdots, A_k$ such that similarity among nodes (i.e. frames) within each set are maximized, and while minimizing the similarities of nodes between sets. To achieve this, \cite{mincut1997} proposed the following objective:
\begin{equation*}
\text{Ncut}_k(V) = \frac{cut(A_1, V-A_1)}{vol(A_1)} + \cdots + \frac{cut(A_k, V-A_k)}{vol(A_k)}    
\vspace{-1mm}
\end{equation*}
where $cut(A,B) := \sum_{u\in A, v\in B}w(u,v)$, and $vol(A) := \sum_{u\in A, v\in V}w(u,v)$. For sequential data, the above minimization problem can be solved using a dynamic programming algorithm~\cite{mincut2006} 
in $O(KN^2)$ time. Here $K$ is the number of partitions (estimated number of syllables in the utterance in our case), and $N$ is the number of nodes (speech frames). 
$K$ needs to be set up-front for every utterance, and we use a hyperparameter second-per-syllable (secPerSyllable) to decide $K$ based on the duration of the utterance. 
In practice, we use the variant introduced in~\cite{minCutMerge2012}, where we first oversegment featSSM, and then iteratively merge temporally adjacent partitions if the cosine similarity of the averaged features belonging to the two partitions falls below some threshold (denoted as mergeThres). We found that this variant always outperformed the original algorithm proposed in~\cite{mincut2006}.
 
\textbf{Clustering.} With hypothesized syllabic segment boundaries produced by the min-cut algorithm, we further use a 2-step clustering approach to categorize the segments. Average features within each segment are used as the embedding of the segment. We initially cluster the segment embeddings using KMeans to produce a large number of clusters, and then run agglomerate clustering to merge similar clusters.
We found our 2-step clustering approach to work better compared to just using Kmeans, given the same number of final clusters. Since our work and~\cite{Peng2022WordDI} are both based on VG-HuBERT, we denote \cite{Peng2022WordDI}'s segmentation approach as \textbf{$\text{VG-HuBERT}_{\text{cls}}$}, where the CLS attention is used to segment speech, and denote our approach as \textbf{$\text{VG-HuBERT}_{\text{featSSM}}$}, where the min-cut algorithm is used on featSSM for segmentation. Both approaches used the 2-step clustering method for segment categorization.

\section{Experiments}
\subsection{Datasets}
Following~\cite{Peng2022WordDI}, the training dataset is SpokenCOCO~\cite{Hsu2020TextFreeIS}, an image-English spoken caption dataset built on top of the MSCOCO image-text caption dataset~\cite{Lin2014MicrosoftCC}. For evaluation on English, we use the test set of SpokenCOCO. Since SpokenCOCO does not have syllable alignment, we first use the Montreal Forced Aligner\footnote{https://montreal-forced-aligner.readthedocs.io/en/latest/} to generate phonetic and word alignment, and then derive the corresponding syllable alignment utilizing a rule-based syllabification script\footnote{https://github.com/kylebgorman/syllabify}. For cross-lingual generalization experiments, we follow~\cite{rasanen18} and evaluate our approaches on Estonian syllabic segmentation using the Phonetic Corpus of Estonian Spontaneous Speech~\cite{lippus2013phonetic}, which contains conversational speech between two test subjects recorded with near-field microphones. The corpus comes with manually verified syllable transcription and alignment. We also evaluate our approach on the Zerospeech word segmentation task, which contains five languages: Mandarin, English, French, German, and Wolof.

\subsection{Implementation details}
\textbf{Model training.} We use the official open-sourced codebase and training recipe released by Peng and Harwath~\cite{Peng2022WordDI} and train a VG-HuBERT on SpokenCOCO. Model snapshots are saved during training for syllable and word discovery analysis. 

\textbf{Evaluation.} To evaluate segmentation performance, we use precision, recall, F1 and R-value ~\cite{rvalue,Kamper2022WordSO}. For the calculation of above metrics, we use a tolerance window of $50$ms for SpokenCOCO and Estonian following~\cite{rasanen18}, and $30$ms for the Zerospeech Challenge~\cite{Dunbar2022SelfSupervisedLL}.
To evaluate the quality of our syllable clustering, we first match hypothesized syllable segments with the ground truth segments for each utterance. To do so, we use a Hungarian matching algorithm where each segment is a node and edge weights are defined by temporal intersection-over-union between each hypothesized segment and ground truth segment (unmatched segments are assigned to a dummy segment). Then, we follow~\cite{Peng2022WordDI} and use cluster purity and number of detected syllables (DS). A syllable is defined as being detected if it achieves an F1 score greater than $0.5$ for some cluster~\cite{Peng2022WordDI}. To avoid conflating word detection and syllable detection, we only evaluate on multisyllabic words.

\textbf{Hyperparameter tuning.} For SpokenCOCO, we tune the mergeThres to maximize the segmentation R-value on the SpokenCOCO validation set. The number of clusters in Kmeans and agglomerative clustering are fixed at $16384$ and $4096$. For syllabic segmentation on Estonian, we tune the hyperparameters on a validation set created following the procedure introduced in~\cite{rasanen18}, using a subset of the original Estonain corpus~\cite{lippus2013phonetic}. For cross-lingual word segmentation on the Zerospeech challenge, we use the hyperparameters selected from the SpokenCOCO validation set.

\subsection{When do syllables and words emerge during training?}
\begin{figure}
  \centering
      \includegraphics[width=1\columnwidth]{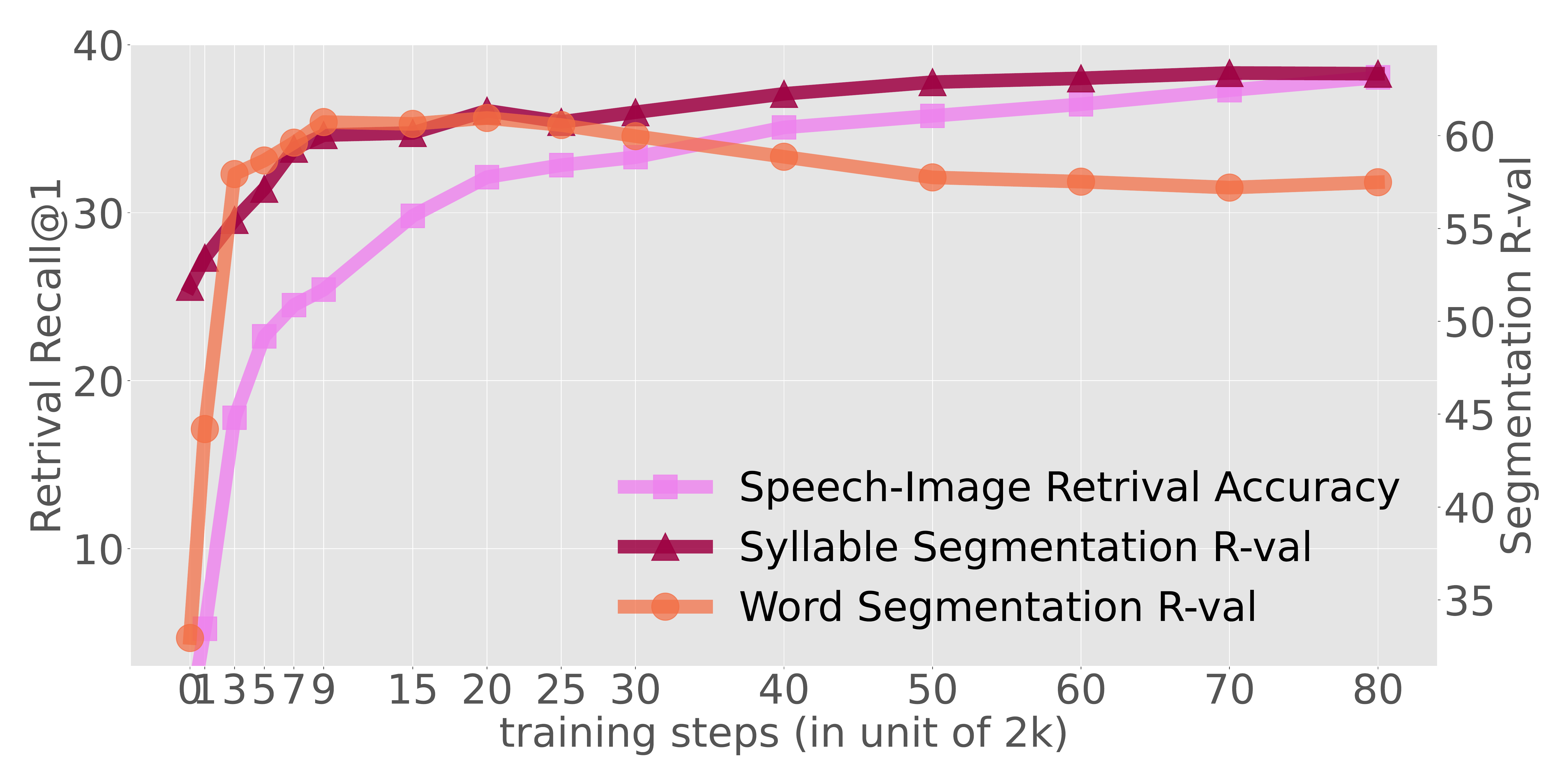} 
      \vspace{-5mm}
      \caption{The performance of speech-image retrieval, and syllable and word segmentation of VG-HuBERT as training progress.}\label{fig:dynamics}
      \vspace{-6mm}
\end{figure}
We first investigate when syllable and word information emerges during the training of VG-HuBERT. In Figure~\ref{fig:dynamics}, we show the syllable and word segmentation performance of VG-HuBERT as a function of training iteration, along with speech-image retrieval accuracy on the SpokenCOCO validation set. Since the contrastive training loss is a direct approximation of the retrieval metric, speech-image retrieval accuracy keeps improving throughout the course of training as expected. For syllabic segmentation, VG-HuBERT reaches the first peak at 20*2k steps, and the performance keeps improving shortly afterwards, with a trend similar to retrieval performance. Interestingly, VG-HuBERT peaks at 20*2k steps for word segmentation, and the performance slightly decreases before levelling off. Anecdotally, by manually examining some examples we found that VG-HuBERT's CLS token tends to ignore more words in the later stages of training. This might be because the model is starting to ignore non-salient words in order to produce semantic representations that are more discriminative in terms of retrieval performance. Notably, as we can see in Figure~\ref{fig:minCut_clsAttn}, syllabic information for the entire utterance tends to persist in the model's representations even when some segments are ignored by the CLS token's attention.

\subsection{Where in the model do syllables and words emerge?}
\begin{figure}
  \centering
      \includegraphics[width=1\columnwidth]{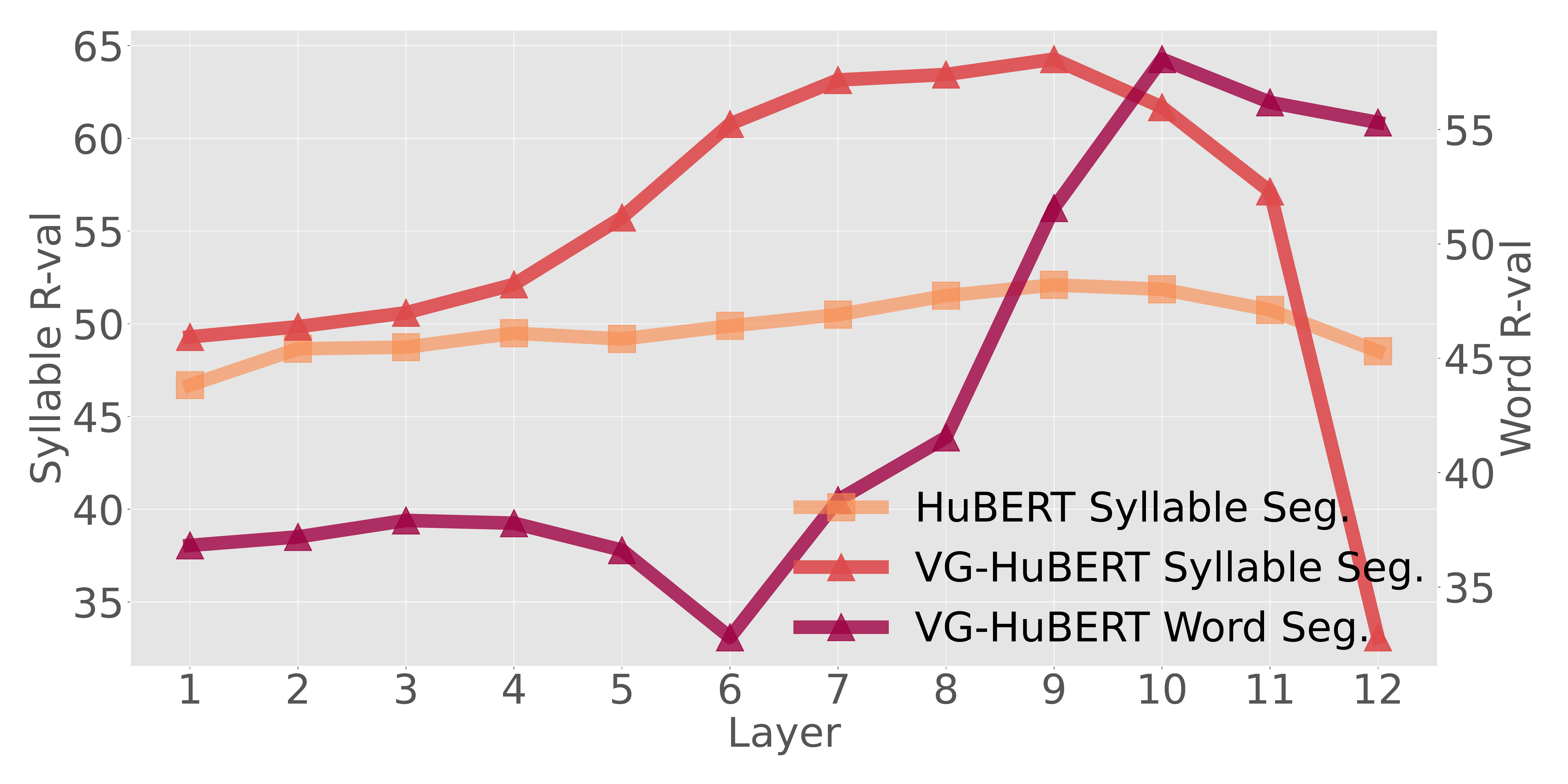}
      \vspace{-6mm}
      \caption{Layer-wise performance of VG-HuBERT on syllable and word segmentation, and HuBERT on syllabic segmentation on SpokenCOCO val set. HuBERT word segmentation gives very poor results~\cite{Peng2022WordDI} and therefore is not shown.}
      \label{fig:hierarchical}
      \vspace{-6mm}
\end{figure}

We next perform a layer-wise study to show how visual grounding helps the emergence of syllables and words, and the interplay between the discovery of different linguistic units. Figure~\ref{fig:hierarchical} compares VG-HuBERT to HuBERT for syllabic segmentation, and also shows VG-HuBERT's word segmentation on the SpokenCOCO validation set. HuBERT performs quite evenly across all layers, while syllabic segmentation is best in VG-HuBERT's mid to late layers, and VG-HuBERT's word segmentation ability is concentrated in the final few layers. We also fine-tuned HuBERT on the SpokenCOCO utterances using its original self-supervised loss to mitigate the potential domain gap, but did not see any improvement in syllabic segmentation (see first two rows in Table~\ref{tab:sc_test}). We see a `division of labor' between different layers in VG-HuBERT with middle layers performing best in syllabic segmentation, while the last three layers specialize in word segmentation. In addition, we note that the best syllabic segmentation layer (layer $9$) is right before the best word segmentation layer (layer $10$), indicating that the attention heads may be learning to string syllables together into words. We leave a more in-depth investigation of this phenomenon for future work. 

\subsection{Syllable discovery on English}
Table~\ref{tab:sc_test} compares VG-HuBERT with other models for syllable discovery on the SpokenCOCO test set. We see that HuBERT performs the worst on this dataset, no matter whether it is fine-tuned on SpokenCOCO or not. $\text{VG-HuBERT}_{\text{cls}}$ denotes the CLS token's attention-based segmentation, a method that has been shown to achieve SotA on word segmentation~\cite{Peng2022WordDI}, gives high precision and low recall on this syllabic segmentation task as expected. In terms of syllable detection, we see that $\text{VG-HuBERT}_{\text{cls}}$ can detect more than $700$ syllables with a high cluster purity. Considering the high cluster purity and low boundary recall of $\text{VG-HuBERT}_{\text{cls}}$, we conclude that this approach is able to discover a smaller number of syllables, but is highly confident of the ones that it does discover.
Oscillator~\cite{rasanen18} is a signal processing-based syllabic segmentation algorithm that achieves SotA for unsupervised syllabic segmentation on multiple languages, including English. Oscillator performs reasonably well on this dataset, only lagging behind our approach on segmentation. Our $\text{VG-HuBERT}_{\text{featSSM}}$ model achieves the best performance in both syllabic segmentation (best F1 and R-val) and clustering (best DS).
\begin{table}
\captionof{table}{Syllabic segmentation performance of different models on SpokenCOCO test set. DS denotes detected syllables.}\label{tab:sc_test}
\vspace{-6mm}
    \begin{center}
        \resizebox{1\columnwidth}{!}{%
    \begin{tabular}{lccccccc}
    \toprule
    Model & Prec. & Rec.  & F1 & R-val.& Purity & DS \\
    \midrule
    HuBERT ft.~\cite{hsu2021hubert} &43.8&49.4&46.4&51.5&29.0&519 \\
    HuBERT~\cite{hsu2021hubert} & 43.8&46.5 &45.1&52.0&30.1&522 \\
    $\text{VG-HuBERT}_{\text{cls}}$~\cite{Peng2022WordDI}&58.7&37.1&45.5&54.3&66.1 & 751\\
    Oscillator~\cite{rasanen18} & 52.0 & 64.6&57.6&57.4 & - & -\\
    $\text{VG-HuBERT}_{\text{featSSM}}$ &57.4&63.6&\textbf{60.3}&\textbf{64.3} & 45.8 & \textbf{902}\\
    \bottomrule
    \end{tabular}}
    \end{center}
    \vspace{-.7cm}
\end{table}

\subsection{Zero-shot syllabic segmentation on Estonian}
Syllables are strongly correlated with speech intensity and voicing, and are organized around sonorant speech sounds~\cite{rasanen18}. This suggests that a syllable detection model trained on one language may able to generalize to other languages. We thus evaluate our English-trained models on a non-English language, namely Estonian. We use the same five-hour subset and evaluation pipeline as~\cite{rasanen18}. Table~\ref{tab:estonian} lists the results. We see that compared to other methods including the Oscillator, our VG-HuBERT performs the best in both F1 and R-val metrics, indicating that its syllabic segmentation ability is at least somewhat language-agnostic.

\begin{table}[!ht]
\caption{Syllabic segmentation on the Estonian corpus.}
\label{tab:estonian}
\vspace{-6mm}
\begin{center}
\resizebox{\columnwidth}{!}{%
    \begin{tabular}{lcccc}
    \toprule
    Approach & Prec.  &  Rec. & F1 & R-val. \\
    \midrule
    $\text{VG-HuBERT}_{\text{cls}}$~\cite{Peng2022WordDI} & 56&77&65&57\\
    HuBERT~\cite{hsu2021hubert} & 64 & 75 & 69 & 70 \\
    WN~\cite{rasanen18}&77&62&69&72\\
    EnvMin~\cite{Wang2007RobustSR}&67&71&69&73\\
    Vseg~\cite{villing2004automatic}&82&63&71&73\\
    Oscillator~\cite{rasanen18}&71&78&74&77\\
    Oscillator (our reprod.) & 72&78&75&78\\
    $\text{VG-HuBERT}_{\text{featSSM}}$  &77&80&\textbf{79}&\textbf{82}\\
    \bottomrule
    \end{tabular}
}
\end{center}
\vspace{-.9cm}
\end{table}

\subsection{Zero-shot word segmentation on unseen languages}
\begin{figure}
  \centering
      \includegraphics[width=1\columnwidth]{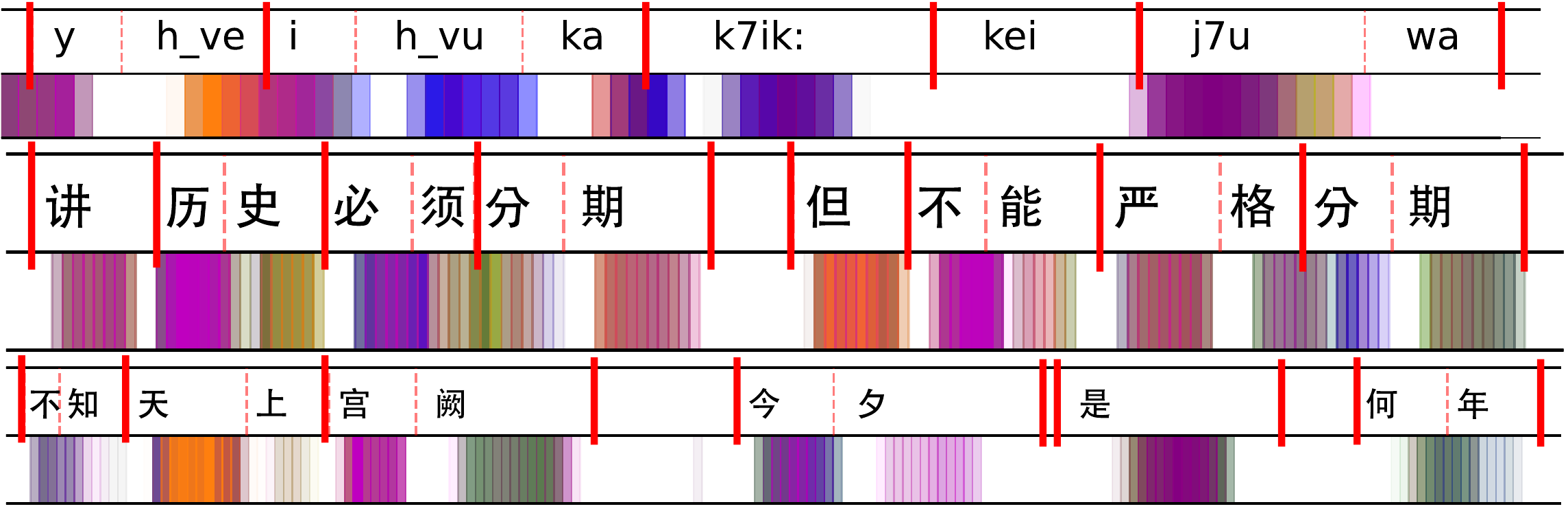} 
      \vspace{-3mm}
      \caption{Visualizations of VG-HuBERT's CLS attention on unseen languages - Estonian and Mandarin. Thin dashed lines denote syllable boundaries, thick vertical line denotes word boundaries. Word boundaries are also syllable boundaries.}\label{fig:estonian_mandarin}
      \vspace{-5mm}
\end{figure}

Lastly, we ask the question: if VG-HuBERT's CLS token detects words in English, what does it do for a language it has not seen during training? To investigate CLS token's behavior on languages unseen during training, we first visualize the CLS attention for Estonian and Mandarin utterances in figure~\ref{fig:estonian_mandarin}. We see that anecdotally, the CLS attention appears to be performing syllabic segmentation, but it sometimes also connect adjacent syllables together. In some cases, the connections give invalid words - in figure~\ref{fig:estonian_mandarin}, for Estonian (the upper figure), `h\_ve' and `i' are connected, but the result is not a valid word; for Mandarin, `必须分' is connected (in the middle figure), and the result is also not a valid word. However, in some other cases, the connections happen to give valid words - in the two Mandarin examples in figure~\ref{fig:estonian_mandarin}, `历史' and `不知' got connected, and they are valid words. 

Based on the observation that the CLS token produces a mixture of monosyllablic and multisyllabic segmentation, we test $\text{VG-HuBERT}_{\text{cls}}$ for word segmentation on the Zerospeech challenge. In table~\ref{tab:zs21}, we see that VG-HuBERT achieves SotA performance on three out of five languages, despite only being trained on English. Interestingly, VG-HuBERT performs very differently on Mandarin and Wolof. While this could be due to hyperparameter settings (we use the same hyperparameters for all languages), we are not able to verify because the Wolof transcripts are not publicly available.

\begin{table}[!ht]
\caption{Word segmentation performance on the Zerospeech Challenge. Token F1 is a stricter metric than boundary F1 where a word is considered a hit only when both it's start and end boundaries are successfully predicted.}\label{tab:zs21}
\vspace{-8mm}
\begin{center}
\resizebox{\columnwidth}{!}{%
    \begin{tabular}{lccccc}
    \toprule
    Approach & Mand.  &  French & Engl. & German & Wolof \\
    \midrule
    PDTW~\cite{Rsnen2020UnsupervisedDO}&4.4 &5.1& 4.1& 2.9& 4.2\\
    ES-KMeans~\cite{Kamper2017AnES} & 8.1&6.3&19.2&14.5&10.9 \\
    SEA~\cite{Bhati2020SelfExpressingAF} & 12.1 & 6.3 & 6.6&6.3&12.6\\
    DP-Parse~\cite{Algayres2022DPParseFW} & 16.0 & \underline{15.3}&\underline{21.9}&\underline{13.4}&\textbf{17.5}\\
    DPDP~\cite{Kamper2022WordSO} & \textbf{26.3} &12.2 & 19.2 & 9.0 & \underline{15.0} \\
    $\text{VG-HuBERT}_{\text{cls}}$  & \underline{19.5} & \textbf{15.5} & \textbf{26.6} &\textbf{15.8}&7.1\\
    \bottomrule
    \end{tabular}
}
\vspace{-.9cm}
\end{center}
\end{table}

\section{Concluding Discussion}
In this paper, we demonstrated that the VG-HuBERT visually-grounded speech model exhibits emergent syllable recognition behavior. We proposed the use of a minimum cut algorithm to automatically extract syllable boundaries from the model's learned representations, and showed that this segmentation ability could transfer to Estonian speech even though the model was only trained on English. Furthermore, we demonstrated that the emergent word discovery ability that is also present in the model could be applied in a zero-shot transfer fashion to segment words in non-English languages, achieving state-of-the-art segmentation performance for several languages in the Zerospeech Challenge benchmark. In our future work, we plan to apply our syllable discovery method to tokenize speech waveforms and use these tokenizations in various textless speech processing tasks such as spoken language modeling and speech-to-speech translation, as well as unsupervised speech recognition.

\bibliographystyle{IEEEtran}
\tiny
\bibliography{mybib}
\end{CJK*}
\end{document}